\documentclass[12pt, letterpaper]{article}
\usepackage{color}

\usepackage{amsmath, amssymb, graphics, epsfig, graphicx, epstopdf}
\usepackage{amsfonts}
\usepackage{amssymb}
\usepackage{graphicx}
\usepackage{amstext}

\newcommand{\nc}{\newcommand}
\nc{\ba}{\begin{eqnarray}}
\nc{\ea}{\end{eqnarray}}

\nc{\be}{\begin{eqnarray}}
\nc{\ee}{\end{eqnarray}}

\nc{\bfk}{{\bf k }}
\nc{\bfx}{{\bf x }}
\nc{\pfp}{{\bf{p}}}
\nc{\bfp}{{\bf{p}}}
\nc{\bfq}{{\bf{q}}}
\nc{\tbf}{\textbf}
\nc{\de}{{\rm d}}
\nc{\calP}{  { \cal P} }
\nc{\calR}{  { \cal R} }
\nc{\im}{ \mathrm{Im} }
\nc{\sg}{ \mathrm{sgn} }
\nc{\smm}{{{}_-}}
\nc{\spp}{{{}_+}}
\def\d{\nabla}
\begin{document}

\vspace{5mm}
\vspace{0.5cm}

\begin{center}

\def\thefootnote{\fnsymbol{footnote}}

{\bf \large Vacuum decay in the presence of a cosmic string}
\\[0.7cm]

{  Hassan Firouzjahi$^{1,2}$ \footnote{firouz@ipm.ir},  Asieh Karami $^{1}$ \footnote{karami@ipm.ir}, Tahereh Rostami$^{1}$ \footnote{t.rostami@ipm.ir}  }
\\ \vspace{0.5cm}

{\small \textit{$^1$ School of Astronomy, Institute for Research in Fundamental Sciences (IPM) \\ P.~O.~Box 19395-5531, Tehran, Iran
}}\\

\vspace{0.2cm}

{\small \textit{$^2$ Department of Physics, Faculty of Basic Sciences, University of Mazandaran, \\ 
P. O. Box 47416-95447, Babolsar, Iran}}\\

\vspace{1cm}

\end{center}

\begin{abstract}

We study the false vacuum decay and bubble nucleation in the presence of a cosmic string in dS spacetime.   
A cosmic string induces a deficit angle in spacetime around itself so the nucleated bubble has the shape of a rugby ball. Working in thin wall  approximation and using the junction conditions
we study the dynamics of the bubble wall  and calculate the Euclidean action.   An interesting feature in our analysis is that the tension of the string is screened by the bubble such that an observer inside the bubble measures a different value of the tension than an outside observer. We show that  the string can act as a catalyzer in which the nucleation rate is enhanced compared to the Coleman-de Luccia instanton. However, in general, the nucleation rate is not a monotonic function of the difference between the two tensions so in some regions of the parameters space the nucleation rate may be  smaller than the Coleman-de Luccia bubble.

\end{abstract}

\vspace{.8cm}

\hrule \vspace{0.5cm}

\newpage

\section{Introduction}

Vacuum decay, the transition from a metastable  false vacuum at a higher energy 
to the stable state of the true vacuum at a lower energy,  is a quantum mechanical phenomenon which happens in various ways.
For a system being initially in its false vacuum, decay happens through quantum tunnelling.  
On the other hand, with a sufficient amount of energy, as in a thermal equilibrium state, it is possible that the system moves over the barrier classically. Furthermore, vacuum decay may happen via quantum tunneling from an excited initial state. The probabilistic nature of quantum processes makes the false vacuum decay happening locally in a random location and time within an initially small volume. Thereafter, a small bubble of true vacuum is nucleated which starts to expand, reaching the speed of light asymptotically  \cite{Coleman:1977py, Weinberg:2012pjx}.  There are signs, after the discovery of the Higgs boson \cite{Aad:2012tfa,CMS:2012nga}, which predict that the Universe's present vacuum state is a false vacuum, so that there exists a true vacuum with lower energy which allows the electroweak vacuum decay to \cite{Degrassi:2012ry,Buttazzo:2013uya}. This is a known concept in Standard Model \cite{Hung:1979dn,Sher:1988mj}. So if the Universe is in a metastable state, quantum processes could occur at any time. Luckily, the lifetime of our vacuum exceeds the cosmic age \cite{Chigusa:2017dux,Chigusa:2018uuj}.

In the treatment of Coleman et al.  an $O(4)$ symmetric bounce solution was considered \cite{Coleman:1977py,Callan:1977pt}
as it was proven that an $O(4)$ symmetric bounce has the least Euclidean action \cite{Coleman:1977th}. In addition,  Coleman and De Luccia (CDL) investigated decay rate in the presence of gravity  \cite{Coleman:1980aw}. From previous intuitions the symmetries are considered to be the same as in the absence of gravity, though this assumption has not been proven in the presence of gravity.  Because of the nonlinear nature of gravity, false vacuum bubbles can form and expand in spacetime.  The created bubbles can have interesting implications in cosmology.\footnote{The interior region of the bubble formed via the CDL instanton is an open Universe. For more reviews, the interested reader is referred to \cite{Turner:1992tz, Bucher:1994gb, GarciaBellido:1995rv, Linde:1999wv, Gong:2008ni, Yamauchi:2011qq, Sugimura:2011tk, Sugimura:2012kr, White:2014aua}.} Intuitively, one may view this as a false-vacuum de Sitter (dS) bubble joined to a surrounding  asymptotically flat (Schwarzschild) spacetime (the true vacuum) by a thin wall. The classical dynamics of false-vacuum bubbles are first  studied in \cite{Sato:1981bf,Kodama:1982sf} 
and further investigated in  \cite{Blau:1986cw} using Israel junction condition \cite{Israel:1966rt}. See also 
\cite{Salehian:2018yoq} for a recent study of bubble formation in $f(R)$ modified gravity. 

It has been shown that bubble nucleation rate in the presence of an inhomogeneity i.e. a black hole (BH) and a topological defect is enhanced in comparison with the CDL instanton \cite{Hiscock:1987hn, Hiscock:1995ma}.  In more detailed analysis \cite{Gregory:2013hja, Burda:2015yfa}, it was shown that the black hole at the center of a bubble acts as a catalyzer for its nucleation and, taking into account the contribution of its conical singularity, the effect is even larger than estimated previously. 
Moreover, it was suggested that compact objects without BH entropies remarkably enhance the nucleation rate \cite{Oshita:2018ptr}. Also in \cite{Oshita:2019jan}, decay rate around a spinning BH was considered and it was shown that the spin of BH would have a suppressing effect. 

With the above discussions in mind, in this work we study the effects of cosmic strings  on nucleation rate in more details. In this picture the bubble is formed symmetrically around the cosmic string which has a negligible thickness. The nucleated bubble in the presence of the cosmic string has the $O(1)$ symmetry rather than the usual $O(4)$ symmetry of the CDL bubble. In \cite{ Hiscock:1995ma}, Hiscock calculated the Euclidean action for nucleating bubble around a cosmic string in flat spacetime  
in thin wall approximations. It was shown that the action, being proportional to the deficit angle of the cosmic string,  is less than the $O(4)$ action in flat spacetime in the absence of cosmic string.   Herein, we  
consider the effects of cosmic strings on vacuum decay with the effects of gravity included. We  use the Israel junction condition (IJC) to find dynamics of the bubble wall in a dS Universe containing a cosmic string.  We calculate the Euclidean action for the nucleation rate and show that it is different than that of CDL  $O(4)$ symmetric bounce, with an amount  proportional to the deficit angle of the cosmic string.  We assume that  the energy scale of bubble nucleation is much lower than the energy scale of topological defect formation so  the string has no thickness.

\section{Thin wall bubble around cosmic strings}
\label{DR}

A cosmic string is a co-dimension two topological defect which has interesting cosmological effects \cite{Kibble:1976sj}. A network of cosmic string can reach the scaling regime, meaning that the contribution of strings in the total energy density furnish a small sub-dominant portion of the total energy density either in radiation dominated era or in matter dominated era \cite{ Vilenkin:2000jqa}. The interests in cosmic strings and their cosmological implications had a revival of interest in early 2000 due to the realization that cosmic superstrings can form at the end of brane inflation which may provide a valuable link to connect string theory to low energy phenomena \cite{Sarangi:2002yt, Majumdar:2002hy}, for reviews see e.g. 
\cite{HenryTye:2006uv, Kibble:2004hq, ACD, Mairi}.

As mentioned before,  inhomogeneities  and  defects  affect the vacuum decay and nucleation rate. In this work we study in details the effects of cosmic string on vacuum decay and bubble nucleation including the effects of gravity. While it is natural to expect a network of cosmic string in early universe, but to simplify the analysis we consider a single straight string in vacuum in dS spacetime. While this picture may be oversimplified, but one may imagine that during the inflationary expansion the network of strings is diluted so in average one string may remain in a single patch. This is  in line with Guth's original idea as how inflation can solve the problems with the unwanted topological defects \cite{Guth:1980zm}. Even in this simplified picture there are interesting lessons to learn. 

The spacetime around an isolated cosmic string is locally flat but  globally there is a deficit angle around the cosmic string which is proportional to $G \mu$ in which $G$ is the Newton constant and $\mu$ is the tension of the string \cite{Vilenkin:1981zs, Gott:1984ef, Hiscock:1985uc, Linet:1986ba}. Indeed, all gravitational effects of cosmic strings, such as  lensing or generating anisotropic patterns on CMB maps,  
are encoded on the parameter $G \mu$. Observationally, there is the  bound $G \mu \lesssim 10^{-7}$. 

To study the effects of the string on vacuum decay, we consider a string in a dS spacetime.  The vacuum solution of string in dS background in polar coordinate  $(r, \phi, z)$ has the form \cite{Gregory:1988bn, Abbasi}
\be
ds^2 = -dt^2
+e^{2Ht}\Big(dr^2+r^2(1-4\mu G)^2 d\varphi^2+dz^2\Big),
\ee
where $H$ is the Hubble constant. As we see from the above metric the spacetime around the cosmic string is locally flat.  Performing the transformation $\varphi\rightarrow (1-4 \mu G)\varphi$ the metric retains the form of  conformally flat dS spacetime but with a deficit angle $8 \pi G\mu$.  Also in order to prevent singularity, we require that  $G \mu < 1/4$ which will be imposed in our analysis throughout.

\begin{figure}[t]
\center
	\includegraphics[width=0.8\textwidth]{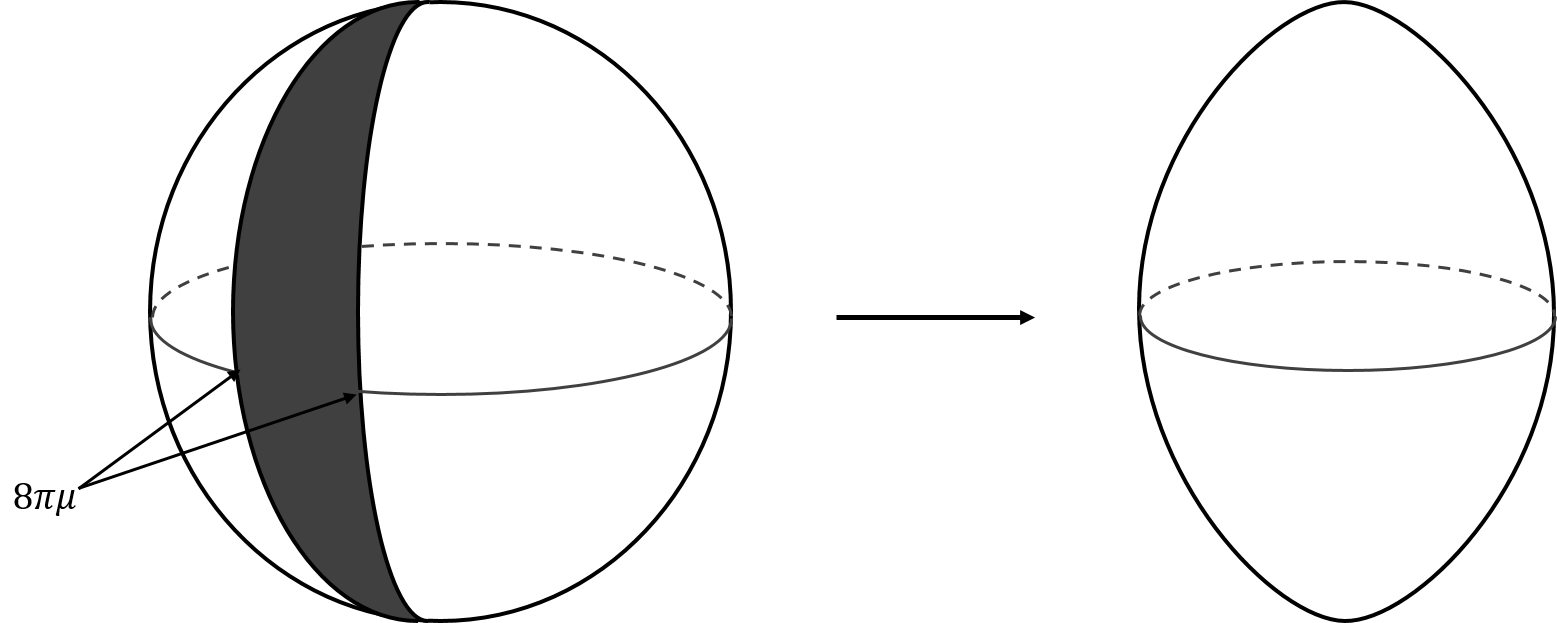}\label{Rug}
	\caption{A $2$-sphere where a wedge of angular width $8\pi\mu$ is removed and the two edges are identified resulting in a rugby ball shape.}
\end{figure}

For our  purpose of bubble formation with an almost spherically symmetric shape, we go to spherical coordinate by making the following transformation
\ba
 \bar{r}=\left(r^2+z^2\right)^{\frac{1}{2}} ,\quad \quad 
\theta=\tan\left(\frac{r}{z}\right)^{-1} \, ,
\ea
which yields
\be
ds^2 = -dt+e^{2Ht}\bigg(d \bar{r}^2+
+ \bar{r}^2d\theta^2+ \bar{r}^2\sin^2\theta(1-4\mu G)^2 d\varphi^2\bigg) \, .
\ee
Now, by introducing the static coordinates  transformation defined as \cite{Ghezelbash:2002cc, BezerradeMello:2009ng} 
\be
t=\sqrt{1-r_s^2 H^2}\sinh(Ht_s),\,\,\,\,\,\,\,\,\,\bar{ r}=\sqrt{1-r_s^2 H^2}\cosh(Ht_s) ,\,\,\,\,\,\,\,\,\
\varphi=\varphi,\,\,\,\,\,\,\,\,\ \theta=\theta,
\ee
 the comic string metric can be recast into static coordinates 
\be\label{ds-cs static}
ds^2 = -f(r_s)dt_s^2+\frac{1}{f(r_s)}dr_s^2+
+r_s^2\bigg(d\theta^2+\sin^2\theta(1-4\mu G)^2 d\varphi^2\bigg),
\ee
with $f(r_s)=1-r_s^2 H^2$ representing the dS metric with deficit angle. Note that there is a cosmological horizon at $r_s=H^{-1}$. Hence, as expected, the effect of cosmic string on dS spacetime is only via the deficit angle. Therefore, in Penrose diagram each point represents a $2$-sphere from which a wedge of angular width $8\pi G\mu$ has been removed and two edges are identified as shown in Fig. \ref {Rug}. This has the shape of a rugby ball  \cite{Carroll:2003db, Aghababaie:2003wz}.

We can now treat the bubble as a rugby ball with a deficit angle of $8\pi G \mu$.
As mentioned before, the bubble wall expands and reaches the speed of light, dividing the spacetime into two regions. The interior true vacuum is separated from the exterior false vacuum region via the bubble wall. Herein, we focus our attention on the dynamics of the expansion of the bubble wall.

The equations of motion of the bubble wall are obtained using the IJC treatment of singular hypersurface.
Briefly, IJC formalism \cite{Israel:1966rt} is about spacetimes with a specified singular hypersurface $\Sigma$. One can foliate the spacetime in such a way that $\Sigma$ is one of the hypersurfaces in the foliation. Therefore, the induced metric $h$ and the extrinsic curvature $K$  of hypersurfaces must have well-defined  limits as one approaches the singular hypersurface $\Sigma$ from each side. Specifically, the induced metric must be continuous across $\Sigma$ while the extrinsic curvature $K_{\mu \nu}$ satisfies the following jump condition  (we choose the units such that $G=1$)
\be
\label{ijc}
\big[K_{ij} \big]=-8\pi \Big(S_{ij}-\frac{1}{2}h_{ij} \mathrm{Tr}[S]\Big),
\ee
where $h_{ij}$ denotes the $3$-metric of $\Sigma$ and  the symbol $[\,  ]$ denotes the discontinuity across $\Sigma$, i.e. $[ X] \equiv X_+ -  X_-$ where the $+$ and $-$ represents the exterior and the interior of $\Sigma$ respectively. Furthermore,  $S_{ij}$ is the surface stress tensor. For our bubble wall, the energy-momentum tensor with surface energy density $\sigma$ is
\be
S_{ij}=-\sigma h_{ij},
\ee

Suppose the spacetime $\mathcal M$ is divided into two regions $\mathcal M^{+}$ and $\mathcal M^{-}$, by a bubble wall $\Sigma$.  We take the line element in $\mathcal M^{+}$ and $\mathcal M^{-}$ to be dS-type with a  cosmic string within. Each region $\mathcal M^{\pm}$ is equipped with spherical coordinates   $\{t_\pm, r_\pm, \theta_\pm , \phi_\pm\}$ with the line element 
\be
ds^2 = -f_+dt_+^2+\frac{dr_+^2}{f_+}+
+r_+^2\left(d\theta_+^2+\sin^2\theta_+\omega_+^2 d\varphi_+^2\right),
\ee
and
\be
ds^2 = -f_-dt_-^2+\frac{dr_-^2}{f_-}+
+r_-^2\left(d\theta_-^2+\sin^2\theta_-\omega_-^2 d\varphi_-^2\right),
\ee
where we have defined 
\be
\omega_\pm\equiv 1-4\mu_\pm \, .
\ee
Note that we have allowed for the possibility that the tension of string as measured by each observer to be different: the interior observer measures it to be $\mu_-$ while for an exterior observer it is measured to be $\mu_+$. 

The bubble wall is a  three dimensional hypersurface and has the metric
\be
ds^2 = -d\tau^2+R^2(\tau)(d\theta^2+\sin^2\theta d\varphi^2),
\ee
where $R(\tau)$ is the radius of the bubble wall measured by an observer on $\Sigma$.
The tangent vector on $\Sigma$ is defined as
\be
\frac{d}{d\tau}=\dot{r}\frac{\partial}{\partial r}+\dot{t}\frac{\partial}{\partial t},
\ee
while the unit normal vector $n_{\mu}$  to $\Sigma$  is given by
\be\label{normal}
n_{\mu}=\left(-\dot{r}_{\pm},\dot{t}_{\pm},0,0\right)\, ,
\ee
where here and below a dot means $d\over d\tau$.

\begin{figure}[t]
\center
	\includegraphics[width=0.5\textwidth]{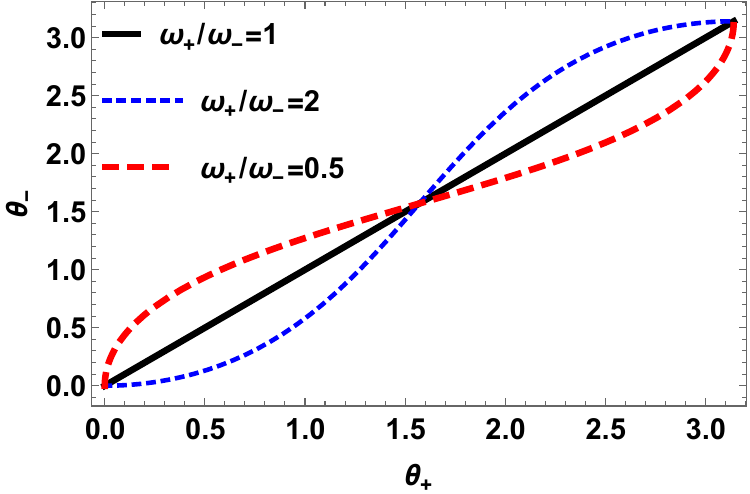}
	\caption{$\theta$ for different values of ${\omega_+/\omega_-}=1, 0.5, 2$. The end points $\{0,\pi\}$ and the equator $\pi/2$ are the same for both spacetimes on two sides of  bubble wall.\label{theta}}
\end{figure}

The conditions of the continuity of the metric are
 \ba
 \label{con0}
 d\tau^2 = f_+dt_+^2-\frac{dr_+^2}{f_+}
 =f_-dt_-^2-\frac{dr_-^2}{f_-}
 \ea
 and
 \ba
 \label{con}
R^2(\tau) d\theta^2=r_\pm ^2d\theta_\pm^2, \quad \quad 
 R^2(\tau)\sin^2\theta d\varphi^2=r_\pm ^2\omega_\pm^2\sin^2\theta_\pm d\varphi_\pm^2,
 \ea
 In addition,  we have $d\varphi_+=d\varphi_-$ which leads us to
 \be\label{con2}
 R^2(\tau)\sin^2\theta=r_\pm ^2\omega_\pm^2\sin^2\theta_\pm \, .
 \ee
Using the continuity conditions (\ref{con}) we obtain 
\be\label{con3}
\theta=2\arctan\Big[\big(\tan\frac{\theta_{\pm}}{2}\big)^{\omega_{\pm}}\Big].
 \ee
 Note that the coordinates $\theta_\pm$ are not  the same in both spacetime. They vary in both spacetimes depending on the value of $\mu_{\pm}$ but with the same values at the end points and in the equatorial circle on the surface of the wall. This is seen in Fig. \ref{theta} for different values of ${\omega_+/\omega_-}=1, 0.5, 2$. As can be seen, their values are different in the range $(0,\pi)$.

Now, let us consider the junction condition Eq. (\ref{ijc}) for the extrinsic curvature. The extrinsic curvature is related to the normal vector via  $k_{\mu\nu}=n_{\mu ;\nu}$. The junction condition yields  
\be\label{Extr}
\big[K_{\theta\theta} \big]=-4\pi\sigma R^2 \, .
\ee
On the other hand, the extrinsic curvature on each side is
\be
K_{\theta\theta}=R\left(\frac{\partial R}{\partial r}n_r-\frac{\partial R}{\partial t}n_t\right)=R f(R)\left(\dot t\frac{\partial R}{\partial r}+\dot r\frac{\partial R}{\partial t}\right),
\ee
 and by using the normalization condition ($f\dot{t}^2-\frac{\dot{r}^2}{f}=1$), we obtain
\be
\label{sigma}
\frac{1}{R} f_+(R)\dot t_+=\frac{1}{R} f_-(R)\dot t_-=-4\pi G \sigma.
\ee
To simplify our calculations we define the following quantity
\be
M(\tau)=4\pi \sigma R(\tau)^2,
\ee
which can be interpreted as the mass of the shell associated  to the surface of  the bubble. 
From the above equation we have
\be
\label{mass}
M(\tau)=-\left(K_{\theta\theta}^+-K_{\theta\theta}^-\right) \, .
\ee
Equation. (\ref{Extr}) can be recast as
\be
\label{mass}
(K_{\theta\theta}^{+})^2=\frac{1}{4M^2}\left[ (K_{\theta\theta}^{+})^2-(K_{\theta\theta}^{-})^2+M^2\right]^2 \, .
\ee

Using 
\be
\dot R= \frac{\rm{d} R}{\rm{d} \tau}=\dot{r}\frac{\partial R}{\partial r}+\dot{t}\frac{\partial R}{\partial t} \, ,
\ee
and  the normalization condition
\be\label{k+}
\big(K_{\theta\theta}^{\pm} \big)^2=R^2f^2\left[(\frac{\partial R}{\partial r})^2-(\frac{\partial R}{\partial t})^2+\dot{R}^2\right],
\ee
we obtain
\be\label{19}
\big(K_{\theta\theta}^{\pm } \big)^2=R^2\left[\dot{R}^2+(1-\frac{2m_{\pm}}{R})\right],
\ee
where we have defined the Misner-Sharp mass \cite{Misner:1964je, McVittie:1933zz}, $m_{\pm}$  as 
\be\label{cmass}
m_{\pm}& \equiv&\frac{R}{2}\left(1-g^{\mu\nu}\partial_{\mu}R\partial_{\nu}R\right) \nonumber\\
&=&\frac{R}{2}\bigg(H_\pm^2R^2-16\mu_\pm^2+8\mu_\pm\bigg).
\ee

Finally from equations (\ref{mass}),(\ref{19}) and (\ref{cmass}) we obtain the equation of motion 
for the evolution of the shell
\be
\label{EOM}
\dot{R}^2&=&\Big(\frac{4R(m_+ -m_-)}{M}\Big)^2+4(m_+ +m_-)+\frac{M^2}{4R^2}-1\nonumber\\
&=&\bigg [\bigg(\frac{H_+^2-H_-^2}{8\pi\sigma}\bigg)^2+\frac{H_+^2+H_-^2}{2}+4\pi^2\sigma^2\bigg]  R^2+\frac{\kappa_1^2}{64\pi^2\sigma^2 R^2}+\frac{\kappa_1(H_+^2-H_-^2)}{32\pi^2\sigma^2}+\frac{\kappa_2}{2}-1,\nonumber\\
\ee
where we have defined the new parameters 
\be
\kappa_1 \equiv 8(\mu_+-\mu_-)(1-2\mu_+-2\mu_-),~~~~~~~~~~\kappa_2 \equiv 8(\mu_++\mu_-)-16(\mu_+^2+\mu_-^2) \, .
\ee

The above equation can be written in the form of a particle moving under a one dimensional potential $U(R)$ via
\be
(\frac{dR}{d\tau})^2+U(R)=0,
\ee
with the effective potential given by 
\be
U(R)=-\alpha R^2-\beta-\frac{\gamma}{R^2},
\ee
where we have defined 
\be
&&\alpha \equiv \Big(\frac{H_+^2-H_-^2}{8\pi\sigma}\Big)^2+\frac{H_+^2+H_-^2}{2}+4\pi^2\sigma^2,\\
&&\beta \equiv \frac{\kappa_1(H_+^2-H_-^2)}{32\pi^2\sigma^2}+\frac{\kappa_2}{2}-1,\\
&&\gamma \equiv \frac{\kappa_1^2}{64\pi^2\sigma^2}.
\ee

The dynamics of the evolution of the wall is controlled by the effective potential $U(R)$. 
To categorize different solutions, we need to find the roots of $U(R)$. The roots are of the form
\be
\label{R2}
R^2=\frac{-\beta\pm\sqrt{\beta^2-4\alpha\gamma}}{2\alpha}.
\ee 
The condition $R^2 \ge 0$ requires that 
\be
\label{sol1}
\beta-2\sqrt{\alpha\gamma}\leq0 ~~~~~~\mathrm{and}~~~~~~\beta+2\sqrt{\alpha\gamma}\leq0
\ee
with the special case of $\gamma=0$ to be treated separately. 

For simplicity, let us define
\be
\epsilon \equiv H_+ - H_- ~~~~~~~~~~\rm{and}~~~~~~~~~~\mu \equiv \mu_+ - \mu_-.
\ee
If $\gamma=0$, $R$ will have one root which is  positive and nonzero. In that case $\mu$ can take either the value  $\mu=0$ or $\mu=(1-4\mu_-)/2$. But since $\mu_+$ cannot be negative the latter case is not acceptable and we are led to $\mu=0$ corresponding to $\mu_-= \mu_+$.   

To find the range of parameters which satisfy the above conditions, we solve the equations $\beta\pm2\sqrt{\alpha\gamma}=0$ and then by using their solutions, we can find the range of $\mu$ in which $\beta\pm2\sqrt{\alpha\gamma}$ are nonpositive. The equivalent forms of $\beta\pm2\sqrt{\alpha\gamma}=0$ are as follows
\be
\label{sol2}
\frac{\kappa_1\xi_-}{32\pi^2\sigma^2}-(1-4\mu_-)^2=0,
\ee
and
\be
\label{sol3}
\frac{\kappa_1\xi_+}{32\pi^2\sigma^2}-(1-4\mu_-)^2=0,
\ee 
 where $\xi_\pm$ are defined as
\be
\xi_\pm=2H_-\epsilon+\epsilon^2+16\pi^2\sigma^2\pm\sqrt{(2H_-\epsilon+\epsilon^2+16\pi^2\sigma^2)^2+64H_-^2\pi^2\sigma^2}.
\ee
It is easy to see that $\xi_-\leq0$ and $\xi_+\geq 0$. 
Correspondingly, the solutions for the Eqs.(\ref{sol2}) and (\ref{sol3}) respectively are
\be
\mu^{(1)}_{\pm}=(\frac{1}{4}- \mu_-)\Big(1\pm\frac{\sqrt{\xi_-(\xi_--1)}}{\xi_-}\Big),
\ee
and
\be
\mu^{(2)}_{\pm}=(\frac{1}{4}- \mu_-)\Big(1\pm\frac{\sqrt{\xi_+(\xi_+-1)}}{\xi_+}\Big).
\ee
One should note that if $\xi_\pm$ are non-zero then we have $\mu^{(1)}_{-}>\mu^{(1)}_{+}$ and $\mu^{(2)}_{-}<\mu^{(2)}_{+}$.

Therefore inequalities Eq. (\ref{sol1}) hold when $\mu^{(1)}_+\leq\mu\leq\mu^{(1)}_-$ and $\mu\leq\mu^{(2)}_-,\ \mu\geq\mu^{(2)}_+$. However,  since $\gamma=0$ is also a solution for both inequalities in Eq. (\ref{sol1}), we can conclude that if $\mu$ is in the interval $[\mu^{(1)}_+,\mu^{(2)}_-]$ or $[\mu^{(1)}_-,\mu^{(2)}_+]$ then  Eq.(\ref{sol1}) is satisfied. However,  we should note that since the tension of string, $\mu_-$, cannot exceed $1/4$ therefore $\mu$ should be less than $(1-\mu_-)/4$; on the other hand $\mu_+$ cannot be negative so $\mu$ should be greater than $-\mu_-$ which means that the only allowed region for $\mu$ is the interval $[\textrm{max}(\mu^{(1)}_+,-\mu_-),\textrm{max}(\mu^{(2)}_-,-\mu_-)]$.

\begin{figure}[h!]
	\includegraphics[width=0.9\textwidth]{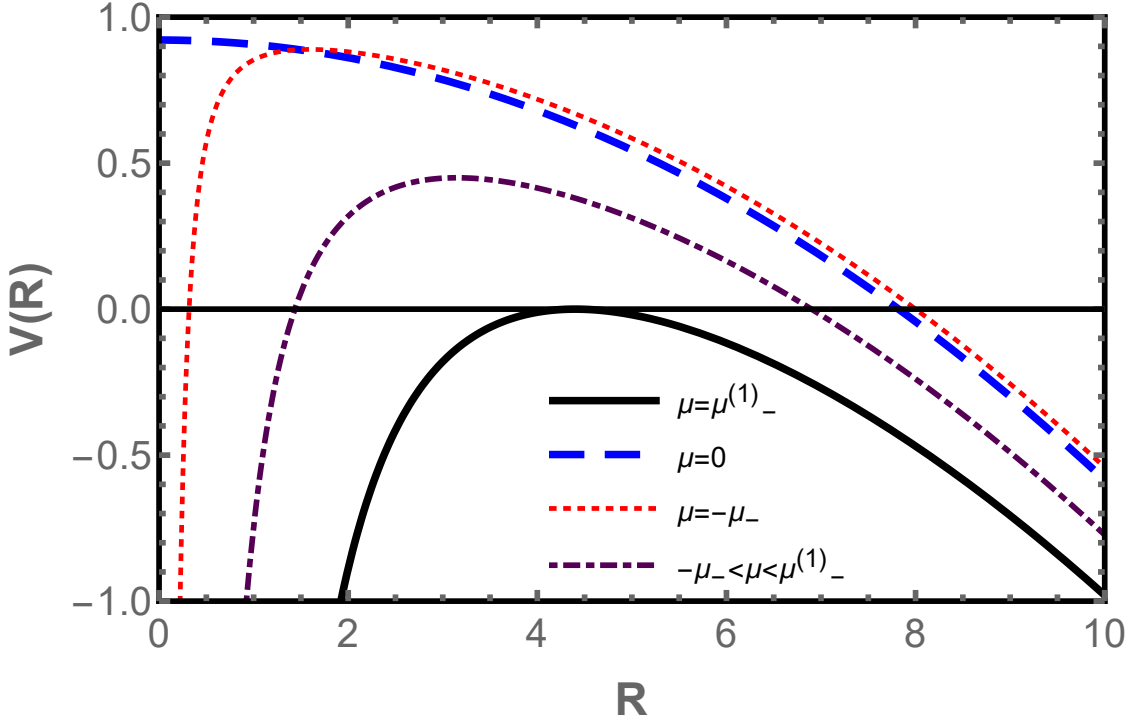}
	\caption{Potential with different values of $\mu$ where $\mu_-=0.01$, $H_-=0.1$, $\epsilon=0.01$ and $\sigma=0.01$. 
	\label{pot-fig}}
\end{figure}

The shape of the potential for various cases of $\mu$ is plotted in Fig. \ref{pot-fig}. 
If $\mu$ is in the open interval $\big( \textrm{max}(\mu^{(1)}_+,-\mu_-),\textrm{max}(\mu^{(2)}_-,-\mu_-) \big)$ then there are two physical solutions for $R$ in Eq.(\ref{R2}), represented by the dashed-dotted (purple) curve in Fig.\ref{pot-fig}, except for the case when $\gamma=0$, which corresponds to the dashed (blue) curve in Fig. \ref{pot-fig}.  In the case when $\mu$ is one of the endpoints then there is just one physical solution for $R$ which makes the potential to be negative or zero, represented by the solid (black) curve in Fig.\ref{pot-fig}. In the first case we have both collapsing and expanding solutions except for the case when $\gamma=0$ where there is just one expanding solution for the evolution of the bubble wall.

\section{Decay rate and Euclidean action}
\label{DR}

Let us now calculate the Euclidean action of our instanton. Analogous to the thin-wall Euclidean bounce solution, we perform a wick rotation $t\rightarrow- i\eta$ . Thus the metric (\ref{ds-cs static}) becomes Euclidean with the coordinate $\eta$ having a period of $\beta_k$. The bubble wall trajectory is now oscillatory $\left(\eta(\tau), r(\tau)\right)$ and in the thin wall limit it is the locus where the exterior of the bubble ($r>r(\tau)$) should be matched with the interior region ($r<r(\tau)$). There is a conical singularity at $r=r_H$ for any arbitrary value of $\beta_k$ where the field equations break down. However, the conical singularity contributes to the Euclidean action as we discuss in the following. Generally speaking,
 the manifold is separated to the right and left of the wall denoted by $\mathcal{M}_{\pm}$ with conical singularities on each side of the bubble wall.

 Hence, the total action splits into four parts
\be
I=I_{\mathcal{H}}+I_{-}+I_{+}+I_{\mathcal{W}},
\ee
where $I_{\mathcal{H}}$ is the contribution from the conical singularity part and  is obtained (the details can be found in Appendix \ref{A-1}) as

\be
 I_{\mathcal{H}}=\frac{-\mathcal{A}_c}{4},
\ee   
in which ${\mathcal{A}_c}$ is the cosmological horizon area
defined as
\be
{\mathcal{A}_c}=\frac{1}{H^2}Vol(S_{2})=\frac{4\pi}{H^2}\left(1-4\mu\right)
\ee
with $Vol(S_{2})$ being the volume of the two-spheres but with the wedge of deficit angle removed. 
 
The action of the thin wall $I_{\mathcal{W}}$ is given by
\be
I_{\mathcal{W}}=\int_{\mathcal{W}} d^3x \sqrt{{}h}\sigma \, ,
\ee
and the remaining bulk actions with Gibbons-Hawking boundary terms are
\be\label{action1}
I_{\pm}=-\frac{1}{16\pi }\int_{\mathcal{M_{\pm}}} d^4x\sqrt{g} \left(\mathcal{R}+16\pi  \mathcal{L}_m\right)+\frac{1}{8\pi G}\int_{\partial\mathcal{M_{\pm}}} d^3x\sqrt{{}^3g}K.
\ee
Note that ${\partial\mathcal{M_{\pm}}}$ denote the  boundaries both on the conical deficit and on the bubble wall. 

To decompose the action into space and time we introduce a family of spacelike surfaces $ \Sigma_t$
labeled by $t$, and a timelike vector field $t_{\mu}$ satisfying $t^{\mu}\d_{\mu}t = 1$ \cite{Hawking:1995fd}. We can decompose $t_{\mu}$ in terms of $u^{\mu}$, the unit normal to the surfaces, into lapse and shift vector as $t^{\mu}=N u^{\mu}+N^{\mu} $.  Using the Gauss-Codazzi equations \cite{Michael Spivak, Shoshichi Kobayashi }, we can decompose the four-dimensional Ricci scalar as follows
\be
\label{G-C}
\mathcal{R}=R-K^2+K_{\mu\nu}^2- 2\d_\mu(u^\mu\d_\nu u^\nu) + 2\d_\nu(u^\mu\d_\mu u^\nu)   \, 
\ee
Substituting this into the action  (\ref{action1}), the two total derivative terms in (\ref{G-C}) results in boundary term contributions. The first term is proportional to $u^{\mu}$ and its contribution is only on the initial and final boundaries which exactly cancel the Gibbons-Hawking boundary terms on these surfaces. However, the second term is orthogonal to $u^{\mu}$  and gives rise to the surface integral on the bubble wall.
Thus, the action (\ref{action1}) takes the form
\be\label{E1}
 I_{\pm} =\frac{1}{ 16 \pi } \int_0^{\beta_k} d\eta  \int_{\Sigma_{\tau}} \sqrt{ {}^3 g}
   ( R + K_{\mu\nu} K^{\mu\nu}
  - K^2 +16\pi  \mathcal{L}_m) \nonumber
\\
- \frac{1}{ 8\pi }\int_{\mathcal{W}} {} K_{\pm}+ \frac{1}{ 8\pi }\int_{\mathcal{W}} n_{\pm\nu}u^{\mu}\d_{\mu}u^{\nu} \, ,
\ee
with $n_{\pm\nu}$ defined in Eq. (\ref{normal}) is  normal to the wall and $\eta$ is the Euclidean time, $\eta = i\tau$.  Note that the integral over $\eta$ is taken over the interval  $0<\eta<\beta_k$. 

Now, let us elaborate more on the first term in Eq. (\ref{action1}). We introduce the canonical momenta $\pi^{\mu\nu}$ and $\pi$ conjugate respectively to $^3 g_{\mu\nu}$ and $\phi$ (the scalar field in the matter Lagrangian). The extrinsic curvature is related to the time derivative of the three-metric $\dot{g}_{ij}$ as
\be
K_{\mu\nu}=\frac{1}{2N}\left[\dot g_{\mu\nu}-2\d_{(\mu}N_{\nu)}\right].
\ee
Hence,  for the first integral we have
\be
 I^{(1)}_{\pm} =\frac{1}{ 16 \pi } \int_0^{\beta} d\eta  \int_{\Sigma_{\tau}} \bigg[{}^3\partial_{\eta}g_{ij}\pi^{ij}-\partial_{\eta}\phi\pi-N\mathcal{H}-N^{i}\mathcal{H}_{i}\bigg],
\ee
where $\mathcal{H}$ and $\mathcal{H}_{i}$ are the Hamiltonian and momentum constraints respectively which both vanish identically $\mathcal{H}=\mathcal{H}_{i}=0$. Furthermore, the static symmetry implies $^3\partial_{\eta}g_{ij}=\partial_{\eta}\phi=0$. 

The second integral  is related to the surface stress tensor $S_{\mu\nu}=-\sigma h_{\mu\nu}$ via $K_+=12\pi \sigma-K_-$. The contribution to the action from the last term of the boundary with $u_{\mu}=\left(\sqrt{1-f},0,0,0\right)$ gives 
\be
u^{\mu}\d_{\mu}u^{\nu}\partial_b=-{\partial_r f\partial_r\over 2},
\ee
 Then the total action for the bubble becomes
\be
I&=&-\frac{1}{4 }\mathcal{A}_c-\frac{1}{2}\int_{\mathcal{W}}d\eta\sigma-\frac{1}{16\pi }\int_{\mathcal{W}}d\eta(\frac{\partial f_+}{\partial r_+}\dot{t}_+-\frac{\partial f_-}{\partial r_-}\dot{t}_-) \, .
\ee

Using Eq. (\ref{sigma}), the action can be further simplified to
\be\label{action Eu}
I&=&\frac{1}{4 }\bigg(-\mathcal{A}_c+2\int_{\mathcal{W}}d\eta R\left(\dot{t}_+-\dot{t}_-\right)\bigg) \, .
\ee
In general, $\dot{t}_{\pm}$ can be positive or negative which correspond respectively to the movement forward and backward in time, i.e. expanding or collapsing bubbles.  Here, we are interested 
in the behaviour of the expanding vacuum bubble around cosmic string, though the probabilities of both solutions are the same. 

In the following first we consider some simple cases where the results can be obtained analytically 
 followed by the general case which requires numerical analysis.

\subsection{Coleman de Luccia bubbles: $\mu_+=\mu_-=0 $ and $\Lambda_-=0$}
\label{CDL}

As a simple check of our formalism, first we  consider the  nucleation of a bubble from dS vacuum to Minkowski vacuum, the CDL process where there is no string  with $\mu_{+}=\mu_-=0$. For this purpose, using Eq. (\ref{EOM}), the Lorentzian radius of the bubble is  obtained to be
\be
R(t)=\frac{1}{\sqrt{\alpha}} \cosh (\sqrt{\alpha} t),
\ee
which in Euclidean time is given by
\be\label{cdl}
R(\eta)=\frac{1}{\sqrt{\alpha}} \cos (\sqrt{\alpha} \eta),~~~~~~~~~ \frac{-\pi}{2\sqrt{\alpha}}<\eta<\frac{\pi}{2\sqrt{\alpha}} \, .
\ee

Wick rotating the time coordinate makes this coordinate to have the period of $\beta_k=\frac{\pi}{\sqrt{\alpha}}$. Thus, the integration in calculating the Euclidean action in equations (\ref{E1})-(\ref{action Eu}) is over one period of the oscillatory Euclidean motion of the wall.  Hence, using Eq. (\ref{cdl}) in Eq. (\ref{action Eu})  with $\mathcal{A}_c={4\pi\over H^2}$ will lead to
\ba
I&=&-\pi\frac{H^2+8 \sigma ^2}{\left(H^2+4 \sigma ^2\right)^2}.
\ea

Then, the exponent of the decay rate $\Gamma\propto e^{-B}$ can be computed using
\ba
\label{B}
B=I-I_{dS} \, ,
\ea
yielding
\be
\label{BCDL}
B_{CDL}&=&\frac{\pi}{H^2}\frac{16 \sigma ^4}{\left(H^2+4 \sigma ^2\right)^2},
\ee
which is the result obtained in \cite{Coleman:1980aw} for the non-singular $O(4)$ instanton.

\subsection{Special case of $\mu_-=\mu_+$ and $H_-=0$}
\label{2.2}

The next simple case which can be solved analytically is when the tensions of the cosmic string in both interior and the exterior of the bubble wall are equal, $\mu_-=\mu_+$, and the transition is  from a dS spacetime to the Minkowski spacetime. In this simplified case we can analytically trace the effects of the cosmic string tension on the nucleation rate.

In this case the radius of bubble in terms of Euclidean time from Eq. (\ref{EOM}) is obtained to be 
\be
R(\eta)={\sqrt{\frac{1-8\mu_-(1-2\mu_-)}{\alpha}}} \cos (\sqrt{\alpha} \eta),~~~~~~~~~ \frac{-\pi}{2\sqrt{\alpha}}<\eta<\frac{\pi}{2\sqrt{\alpha}}\, .
\ee
Since $\mu$ should be less than $1/4$, we conclude that the physical radius of the bubble is always smaller than the CDL one.  However, as we show in the following, the decay rate for this case is greater than that of CDL bubble. 

\begin{figure}[t!]
\begin{center}
	\includegraphics[width=0.7\textwidth]{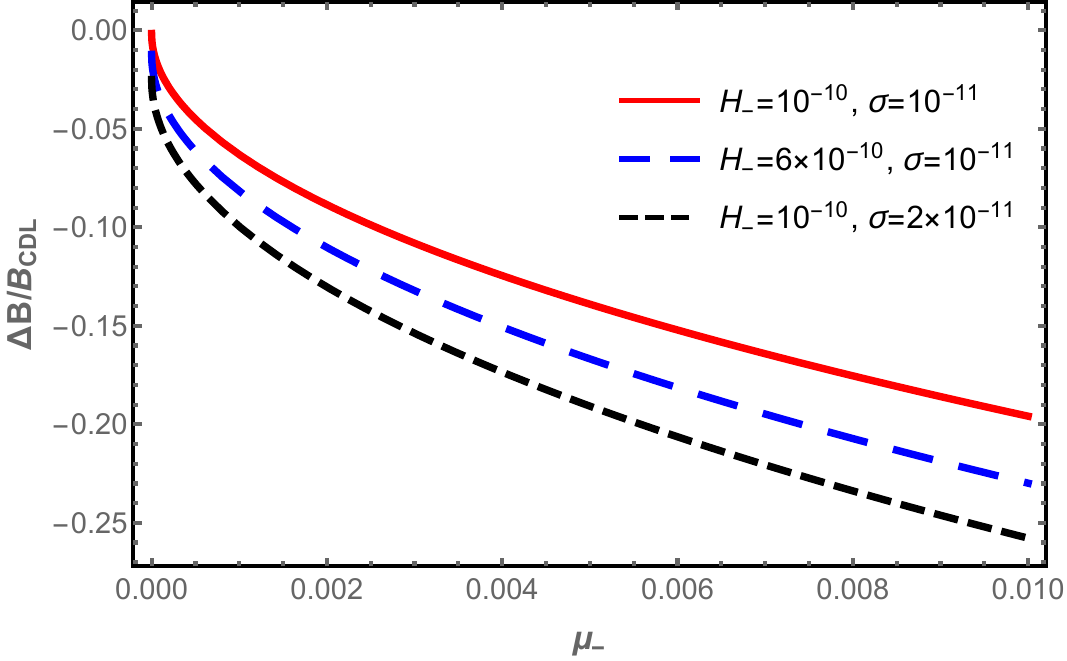}
	\caption{$\Delta B/B_{CDL}$ for different values of $H$ and $\sigma$.\label{DBA}}
	\end{center}
\end{figure}

Herein, the contribution from the conical deficit is given by
\be
I_{\mathcal{H}}=-\frac{\pi}{H^2}(1-4\mu)
\ee
From Eqs. (\ref{action Eu})  and  (\ref{B}), the exponent of the decay rate is
\be
\label{B-case2}
B&=&\Bigg[-\frac{16\omega\sigma^2\sqrt{2\mu_-(1-2\mu_-)}}{(H^2+4\sigma^2)^2}+\frac{16\sigma^4+H^4}{H^2(H^2+4\sigma^2)^2}\tan^{-1}\Big(\frac{\omega}{2\sqrt{2\mu_-(1-2\mu_-)}}\Big)\nonumber\\
&&+\frac{16\sigma^4-H^4}{H^2(H^2+4\sigma^2)^2}\left(\tan^{-1}\Big(2\frac{H^2+4\sigma^2}{H^2-4\sigma^2}\frac{\sqrt{2\mu_-(1-2\mu_-)}}{\omega}\Big)-\frac{\pi}{2}\right)\Bigg].\nonumber\\ 
\ee
It is constructive to compare the decay rate here with the decay rate in CDL, given by  $B_{CDL}$ in Eq. (\ref{BCDL}), so we write $B=B_{CDL}+\Delta B$. 
Subtracting $B_{CDL}$ from Eq. (\ref{B-case2}) yields 
\ba
\Delta B&=&
\frac{16\sigma^4}{H^2(H^2+4\sigma^2)^2}[\tan^{-1}(\frac{\omega}{2\sqrt{2\mu_-(1-2\mu_-)}})-\tan^{-1}(2\frac{H^2+4\sigma^2}{H^2-4\sigma^2}\frac{\sqrt{2\mu_-(1-2\mu_-)}}{\omega})-\frac{\pi}{2}]\nonumber\\
&+&\frac{H^4}{H^2(H^2+4\sigma^2)^2}[\tan^{-1}(\frac{\omega}{2\sqrt{2\mu_-(1-2\mu_-)}})+\tan^{-1}(2\frac{H^2+4\sigma^2}{H^2-4\sigma^2}\frac{\sqrt{2\mu_-(1-2\mu_-)}}{\omega})-\frac{\pi}{2}] \nonumber\\
&-&\frac{16\omega\sigma^2\sqrt{2\mu_-(1-2\mu_-)}}{(H^2+4\sigma^2)^2} \, .
\ea
It can be  shown analytically that $\Delta B \leq 0$ so we conclude that  $B\leq B_{CDL}$; this fact is also supported numerically  in Fig.\ref{DBA}. Consequently, the decay rate is greater than the case of $CDL$ so the string acts as a catalyzer. This is similar to the conclusion in  \cite{Gregory:1988bn} in which it is shown that a black hole acts as a catalyzer for the decay rate.  

\subsection{ General case}
\label{2.3}

The general case of decay from a dS vacuum to another dS vacuum with no restrictions on 
$\mu_\pm$ can not be solved analytically so we have to employ numerical analysis to see the effects of 
cosmic string on the decay rate. 

The radius of the bubble $R(\tau)$ from  Eq. (\ref{EOM}) is obtained to be 
\be
R(\tau)&=&\Big[\frac{\cos \left(2 \sqrt{\alpha } \eta \right)}{8\alpha}\Big(-2 \beta+\left(-4 \alpha  \gamma +\beta ^2+1\right) \\
&&+\sqrt{\left(-4 \alpha  \gamma +\beta ^2-1\right)^2 \sin ^2\left(2 \sqrt{\alpha } \eta \right)+\left(\left(-4 \alpha  \gamma +\beta ^2+1\right) \cos \left(2 \sqrt{\alpha } \eta \right)-2 \beta \right)^2}\, \, \Big)\, \Big]^{1/2}\nonumber
\ee

\begin{figure}[t!]
 \begin{center}
 	\includegraphics[width=0.8\textwidth]{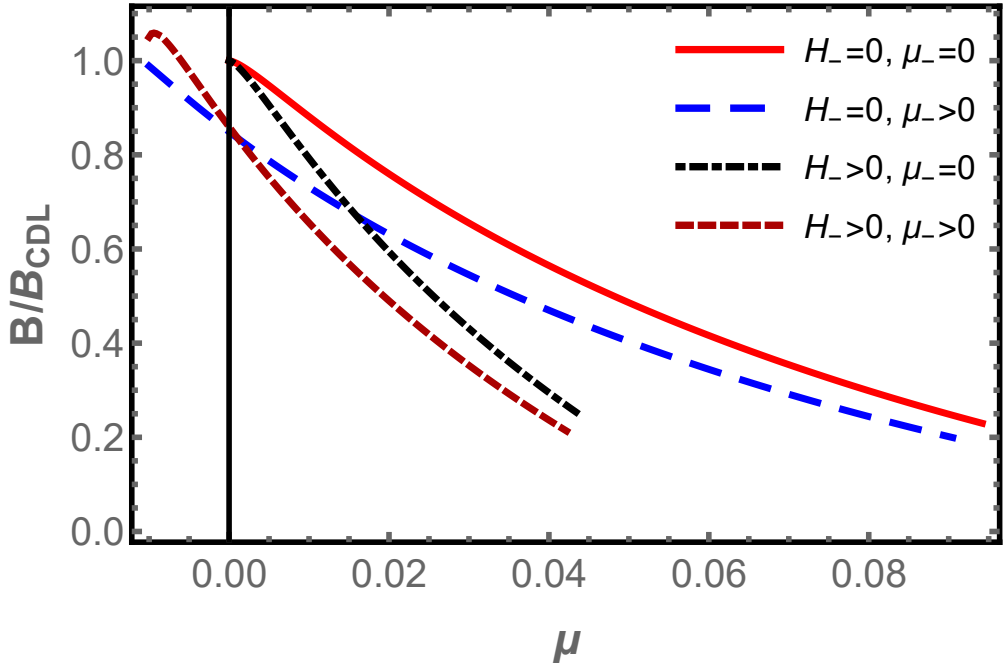}
 	\caption{$B/B_{CDL}$ with different values of $\mu_-$ and $H_-$. 
\label{BvsBcdl-1}}
\end{center}
 \end{figure}

We   solve the Euclidean action numerically and compare it to the CDL action.
 The result is shown in Fig. \ref{BvsBcdl-1}. The starting point from the left of solid thick red line with $H_-=0$ and $\mu_-=0$ represents the value of the CDL tunnelling rate. As the tension of string in false vacuum 
 ($\mu_-$) grows the exponent $B$ falls off and the rate enhances.  
 Moreover, for the case when the nucleation is from dS to dS, shown by the  black dot-dashed curve (with non-zero $H_-$), the nucleation rate is even larger than the CDL tunnelling. In both cases, we have found that for $\mu >0$, as $\mu$  increases, the action reduces compared to CDL and the tunnelling rate increases. In fact, as $\mu_-$ increases the nucleation rate increases too,  indicating that the tension of cosmic string acts as a catalyzer. Interestingly, for a given $\mu_-$,  the  stronger is the screening effect (meaning  that an observer inside the bubble would measure a smaller value of the tension compared to an outside observer),  the higher is the decaying rate.

However, it is important to note that for $\mu<0$ (i.e.  a denser string in true vacuum than in the false vacuum) the decay rate is not a monotonic function of $\mu$ and has a concave shape. Therefore, there are some intervals in which  the string  has a suppressing effect on the decay rate.

\section{Summary}

In this work we have studied the effects of cosmic string on vacuum decay and bubble nucleation. 
Working in  thin wall approximation,  the bubble wall is a timelike surface which divides the spacetime into two regions of the exterior unstable false vacuum and  the interior  stable true vacuum. Using the IJC method the dynamics of the evolution of the bubble wall  is studied and the Euclidean action and the nucleation decay rate are calculated.

 The imprints of the cosmic string on nucleation rate are twofolds. First, the bubble in this case has the shape of a  rugby ball. This is because a cosmic string produces a deficit angle in spacetime around itself in which a wedge of $8\pi\mu$ is removed from the sphere. As a result, the formed bubble is not exactly spherical, but 
 has the shape of a rugby ball.  Second, the tension of the cosmic string and also the difference between the measured tensions of the cosmic string from inside and outside the bubble generally enhance the nucleation rate so, as shown in sections (\ref{2.2}) and (\ref{2.3}),  the string acts as a catalyzer.  

There are  some open questions related to our setup of  bubble nucleation in the presence of cosmic strings which may be studied in future.  An interesting issue is the question of the negative modes of vacuum decay \cite{Lavrelashvili:1985vn, Tanaka:1992zw, Gen:1999gi, Tanaka:1999pj, Lavrelashvili:1999sr}. In \cite{Gregory:2018bdt} it is shown that with a black hole as a bubble nucleation site, there is only one negative mode in contrast to CDL instanton. Therefore, it is worth addressing negative mode in the presence of cosmic string. 
Another direction of work is to look for cosmological imprints of the   bubble nucleation. In \cite{Deng:2017uwc, Deng:2018cxb}, it is shown that bubbles lead to black hole which may provide seeds for primordial black holes as candidates for LIGO observations. This direction can be pursued  in this setup but with the difference that  the bubble is not spherical while there is a string at the core of the bubble which will make the analysis non-trivial. In addition, one may be interested  to investigate the trace of  bubble nucleation around cosmic string during or  after inflation and look for its imprints on CMB anisotropies. To be specific, in \cite{Firouzjahi:2017ssb}  the imprints of CDL instanton during inflation to generate CMB statistical anisotropies and power asymmetry are studied while in  \cite{Jazayeri:2017szw} these analysis were extended to the case of an isolated string during inflation. Combining and extending the methods of  \cite{Firouzjahi:2017ssb, Jazayeri:2017szw} one can look for the effects of our setup of a rugby ball shaped bubble with a string in its core on CMB maps.  Finally,   looking for the direct detection of cosmic strings through gravitational lensing effect \cite{Sazhin:2006kf, Gasparini:2007jj, Morganson:2009yk} can be another direction in which our setup can be searched for observationally. 

\vspace{1cm}

 
 {\bf Acknowledgments:}  A. K. and T. R. would like to thank  Saramadan (Iran Science Elites Federation) for support.


\vspace{0.5cm}

\appendix
\section{Euclidean action and conical singularities }
\label{A-1}

Here we present the details of the analysis for the contribution of the deficit angle to the Euclidean action, similar to analysis of \cite{Gregory:2018bdt}.

The Euclidean metric for the static line element (\ref{ds-cs static}) is regular in the region $r< H^{-1}$. However, this metric has a conical singularity at $r_H=H^{-1}$ \cite{Gregory:2018bdt}. Let us now calculate the contribution of the conical singularity to the action. The  metric in Euclidean time $t\rightarrow -i \eta$  is written in the form
\be
ds^2=f(r_s)d\eta^2+\frac{1}{f(r_s)}dr_s^2+r_s^2d\Omega^2
\ee
which is asymptotically flat and $\eta$ is periodic with period $2\pi\beta$. Thus, the above metric represents a manifold with ${\cal{M}}=\bar{D}\times S^2$, where $\bar{D}$ is the closed two-dimensional disc and $S^2$ is a two-sphere. Near the horizon we define $f(r_s)=f'_H(r-r_H)+{\cal O} (r-r_H)^2$ with $f'_H=f'(r_s)|_{r=r_H}$,  so that  the metric becomes
\be
ds^2=f'_H(r-r_H)d\eta^2+\frac{1}{f'_H(r-r_H)}dr_s^2+r_H^2d\Omega^2 \, .
\ee
Now defining the proper radius distance as
\be
d\zeta=\frac{dr}{f'_H(r-r_H)} \, ,
\ee
the metric takes  the following asymptotic form
\be\label{met cone}
ds^2=\left(\frac{\zeta}{\delta}\right)^2d\eta^2+d\zeta^2+r_H^2d\Omega^2
\ee
with $\delta^{-1}=\frac{f'_H}{2}$. This is the direct sum of a line element of a cone with $\tau$ running from $0$ to $2\pi\delta$. Thus, the metric (\ref{met cone}) near the horizon has the topology of  $C_{\delta}\times S_2$, where $C_{\delta}$ is a cone with deficit angle $2\pi(1-\delta)$. For $\delta\ne 1$ the space is regular everywhere except at $\zeta=0$ where it has a singularity \cite{Hayward:1990tz,Fursaev:1995ef, Frolov:1998wf}.

We can take $\zeta=0$ to be the center of a two-sphere and $\zeta=\varepsilon$ corresponds to the three-surface at the boundary $\partial \cal{M}$. For $0<\zeta\leq\varepsilon$ the coordinate system is regular. Hence we can do the integrals in this range.

In such manifolds with conical deficit, the singularity is smoothed out with a regular function. Let us approximate the cone by a regular function $A(\zeta)$ 
\be
ds^2=A(\zeta)^2d\eta^2+d\zeta^2+r^2d\Omega_H^2,
\ee
such that  $A'(0)=1$ and $A'(\epsilon)=(1-\delta)$, where $2\pi\delta$ is the deficit angle. 

The Ricci scalar in the vicinity of $\zeta=0$ is
\be\label{deficit-R}
{\mathcal{R}}=-\frac{2A''}{A}-\frac{2A'}{Ar}\nonumber\\
\sim -\frac{2A''}{A}+{\cal O}(\zeta) \, .
\ee
As we see, the first term $A''={\cal O}(\frac{(A'(\varepsilon)-A'(0))}{\varepsilon})$ is the unbounded term. For a small region around $\zeta=0$ performing the integration by parts and eliminating the second time derivatives we are thus lead to
\be
\int d^4 x\sqrt{g}{\cal{R}}\sim {\cal{A}}\bigg(A'(0)-A'(\varepsilon)\bigg)+{\cal{O}}(\varepsilon)=4\pi\delta{\cal{A}}+{\cal{O}}(\varepsilon) \, ,
\ee
in which ${\cal{A}}$ is the area of a two-sphere\footnote{For the dS Universe one can see that the area of the cosmological horizon is ${\cal{A}}_c=4\pi r_H^2=\frac{4\pi}{H^2}$}. For the Gibbons-Hawking boundary term, with the normal vector $n=-d\zeta$  and the extrinsic curvature  $K=\frac{A'}{A}-\frac{4}{\zeta}$ we obtain
\be\label{deficit-K}
\int_{\partial{\cal{M}} \,\,\,or  \,\,\ \zeta=\varepsilon} d\tau d\Omega A \zeta^2 K\sim -2\pi{\cal{A}}A'(\varepsilon)+{\cal{O}}(\varepsilon)=-2\pi{\cal{A}}(1-\delta)+{\cal{O}}(\varepsilon) \, ,
\ee

Now, the contribution of the deficit angle, $\beta$ with $\epsilon\rightarrow 0$ is obtained to be
\be
I_{\cal{H}}=-\frac{1}{16\pi G}\int d^4 x\sqrt{g}{\cal{R}}+\frac{1}{8\pi G}\int d^3 x \sqrt{h} K=-\frac{\cal{A}}{4G}=-\frac{\cal{\pi}}{GH^2}\left(1-4\mu\right).
\ee

\end{document}